\documentclass[
    ,final            
  ]
  {aipproc}

\layoutstyle{6x9}

\begin{document}

\title{The BAT-Swift Science Software}

\author{D. M. Palmer}{
  address={Los Alamos National Laboratory}
}

\author{E. Fenimore}{
  address={Los Alamos National Laboratory}
}

\author{M. Galassi}{
  address={Los Alamos National Laboratory}
}

\author{K. Mclean}{
  address={Los Alamos National Laboratory}
}

\author{T. Tavenner}{
  address={Los Alamos National Laboratory}
}

\author{S. Barthelmy}{
  address={Goddard Space Flight Center, NASA}
}

\author{M. Blau}{
  address={Goddard Space Flight Center, NASA}
}

\author{J. Cummings}{
  address={Goddard Space Flight Center, NASA}
}

\author{N. Gehrels}{
  address={Goddard Space Flight Center, NASA}
}

\author{D. Hullinger}{
  address={Goddard Space Flight Center, NASA}
}

\author{H. Krimm}{
  address={Goddard Space Flight Center, NASA}
}

\author{C. Markwardt}{
  address={Goddard Space Flight Center, NASA}
}

\author{R. Mason}{
  address={Goddard Space Flight Center, NASA}
}

\author{J. Ong}{
  address={Goddard Space Flight Center, NASA}
}

\author{J. Polk}{
  address={Goddard Space Flight Center, NASA}
}

\author{A. Parsons}{
  address={Goddard Space Flight Center, NASA}
}

\author{L. Shackelford}{
  address={Goddard Space Flight Center, NASA}
}

\author{J. Tueller}{
  address={Goddard Space Flight Center, NASA}
}

\author{S. Walling}{
  address={Goddard Space Flight Center, NASA}
}

\author{Y. Okada}{
  address={University of Tokyo,  Saitama University and ISAS, Japan}
}

\author{H. Takahashi}{
  address={University of Tokyo,  Saitama University and ISAS, Japan}
}

\author{M. Toshiro}{
  address={University of Tokyo,  Saitama University and ISAS, Japan}
}

\author{M. Suzuki}{
  address={University of Tokyo,  Saitama University and ISAS, Japan}
}

\author{G. Sato}{
  address={University of Tokyo,  Saitama University and ISAS, Japan}
}

\author{T. Takahashi}{
  address={University of Tokyo,  Saitama University and ISAS, Japan}
}

\author{S. Watanabe}{
  address={University of Tokyo,  Saitama University and ISAS, Japan}
}

\begin{abstract}
The BAT instrument tells {\it Swift} where to point to make immediate follow-up observations of GRBs.  The science software on board must efficiently process $\gamma$-ray events coming in at up to 
34~kHz, identify rate increases that could be due to GRBs while disregarding those from known sources, and produce images to accurately and rapidly locate new Gamma-ray sources.

\end{abstract}

\maketitle


\section{Introduction}

BAT is a wide-field coded aperture gamma-ray telescope to be flown on the {\it Swift} Mission[1].
Its main purpose is to detect and accurately locate GRBs so that {\it Swift} may repoint itself to observe the GRBs with its X-ray and UV/Optical telescopes, sometimes while the GRB is still in its $\gamma$-ray flaring phase.
It also provides the direct $\gamma$-ray observations of the GRB in its 15-150 keV energy range.
While waiting for new GRBs, it provides secondary science such as a $\sim$mCrab all-sky survey, high time resolution observations of selected sources, and phase-resolved spectroscopy of pulsars.

BAT has 32,768 CdZnTe detectors, each $4\times 4 \times 2~ {\rm mm}^3$ (5243 ${\rm cm}^2$ total area) and a 2.6 ${\rm m}^2$ mask with a 50\%-filled random pattern of $5 \times 5 \times 1~ {\rm mm}^2$ lead tiles.
The 5 ${\rm mm}$ mask scale divided by the 1 ${\rm m}$ mask-detector separation determines the 17 ${\rm arcmin}$ FWHM angular resolution.
Detected $\gamma$-ray events and other housekeeping information are transmitted by 16 `Spacewire' (IEEE 1355) channels to the primary Image Processor Electronics (IPE) computer and to a redundant cold-spare IPE.

Each IPE includes a 25 MHz Rad6000 that serves as the main processor, and a rad-hard 25 MHz ADSP21020 digital signal processor (DSP) for image production and processing.

The Rad6000 software runs under the VxWorks operating system and is written in C++.  Significant parts of the engineering code ({\it e.g.,} the Command and
Data Handling Software Bus) were originally developed for the Triana spacecraft.  The DSP software is written in C and is based on a simple purpose-built
executive.

This paper briefly describes the instrument-unique science software in the IPE.

\section{Data Ingest}

Each $\gamma$-ray detected is processed by the Data Ingest task.  The digitized signal is converted from ADC units ($\sim 0.2$keV resolution) to equivalent keV by gain and offset parameters continually determined for each detector by an electronic calibration system.
Each event also includes a timestamp with $100 \mu{\rm s}$ resolution referenced to a 1 pulse-per-second hardline signal from the spacecraft.

The data is binned in time, energy, and detector region in multiple ways:
\begin{itemize}
\item[-] On 5 minute time intervals, 80 channel energy histograms are made for each of the 32k detectors.  These are processed on the ground to produce the all-sky survey.
\item[-] Accumulations in 4 energy bins in each of the 32k detectors over 8 seconds are generated and stored to be used in background subtraction when a burst is detected.  
\item[-] The events are pulsar-folded into an 80 channel, 32 phase bin structure using a spin ephemeris based on a $13^{th}$ order polynomial (generated on the ground using the spacecraft's predicted orbit for barycentering).  
\item[-] Sets of 32k detector weights corresponding to the exposure through the mask allow individual `mask-tagged' monitoring [2] of up to 3 selected 
sources.
\item[-] Accumulations in 4 energy bins in the 4 quadrants of the detector are generated every 64 ms and passed to the Trigger Algorithm.  
\item[-] On 5 timescales from 4 ms to 64 ms, special code calculates the count rate in 4 energy ranges and 9 overlapping detector regions (from individual quadrants to the full detector plane) and tells the trigger system the maximum count rate achieved in each of the 180 time-energy-region combinations during a 320 ms time interval.
\end{itemize}

The large collecting area and FOV (1.4 sr to the half-coding point) will produce a nominal count rate that simulations predict to be $\sim 12$ kHz, although this will be greatly exceeded as the spacecraft passes near or through the South Atlantic Anomaly (SAA) in its 22\textdegree\ inclination orbit.  The hardware and software can handle sustained operation at 34~kHz.  Processing at this data rate, using a fraction of the 25 MHz processor, required programming for extreme efficiency (at the cost of clarity and maintainability).

At count rates above 34 kHz, or when the processor has episodic demands placed on it ({\it e.g.}, during data compression for telemetry) a backlog of data
will accumulate in the input buffer. This data is processed in the order it is received so that processing falls behind briefly, to catch up when the CPU
demand eases or the count rate declines.  When the backlog exceeds $5 \times 10^{7}$ counts, it is assumed that we are in a high count rate situation such as
the SAA, and the detector plane is placed in a mode where it merely counts events rather than reporting each one. The detector is returned to normal mode when
the backlog has been processed and the count rate returns to a reasonable level. The SAA threshold is almost two orders of magnitude higher than the expected
highest-fluence burst of the year (equivalent to $\sim 200$ kHz for 5 s), and so we do not expect to lose any data due to excessive flux from a GRB during the
course of the mission.

\section{Triggering}

The triggering code is more completely described elsewhere[3]. The binned count rates produced by Data Ingest allow triggering on timescales from 4 ms to 
many
seconds.  Rate increases are detected by comparing the count rate during a foreground interval to that extrapolated or interpolated from one, two or three
background intervals. Mask-tagging to track the flux of individual sources allows us to ignore count rate increases when they are due to a known source
(vetoing) or to subtract the source's count rate from the time series for triggering (canceling), which mitigates the effects of noisy sources such as Cyg X-1 and Sco X-1.

When the triggering system detects a rate increase, it also determines the optimal energy range and time intervals for foreground and background accumulations.
These time intervals are used to produce a `background-subtracted' detector map, where the pattern of net counts in the detectors as revealed by imaging should be dominated by flux from the varying source.

When no significant rate increase is detected, `image triggering' is attempted by accumulating detector maps on 1 minute, 5 minute, and per-pointing intervals and passing them to the imaging system to detect weaker and slowly varying sources.

\section{Imaging}

Imaging is handled by the DSP coprocessor, which is optimized for the FFT operations that dominate the execution time for imaging.

The detector map is processed to remove variation from systematic effects and known sources in the field of view.  A sequence of balancing steps that subtract the average constant value from all or subclasses of detectors is combined with an optional CLEAN step.

CLEANing uses a linear regression of the detector patterns predicted for multiple bright known sources in the field of view to remove the effects of those sources.  An image made of the residual map is free from the main peaks and systematic noise from those sources, allowing unknown sources to stand out.  The linear regression jointly includes known background variation patterns, such as the tendency for the center of the detector plane to have more solid angle exposure to aperture flux than the periphery.

Because background subtraction eliminates many of the systematic effects, the pre-processing strategies can be separately specified for background-subtracted rate trigger maps and non-background-subtracted image trigger maps.

The pre-processed detector map is then convolved with the mask pattern, using an efficient FFT-based process, to produce what we call an FFT image.  This step can also include Wiener filtering (which has the effect of approximating a numerically stable convolution-deconvolution hybrid) and high-pass filtering with no additional CPU processing time.  The effect on sensitivity of these additional filters will be tested with on-orbit data.

The FFT image has a pixel scale set by the detector spacing (4.2 mm) divided by the 1~m detector-mask spacing, and is comparable to the size of the point spread function.  The signal from a source can thus be split among multiple adjacent pixels.  For this reason the image is searched for peaks consisting of individual, pairs, and sets of 4 pixels, to a level that picks up some noise peaks in addition to true peaks from weak sources.  The peak locations are compared to an on-board catalog to identify known sources, and to ignore them unless they exceed intensities at which they are considered interesting.

Unidentified candidate peaks from the FFT image are used as locations for back-projection (BP) images.  These BP images of small regions of the image plane have fine pixels, allowing the precise locations and intensities of the peaks to be determined.

Roughly 6 seconds after the start of imaging, a list of possible sources in the field of view is available with accurate locations (<4  arcmin), intensity, and detection significance.  If the significance of a candidate source exceeds a statistical threshold, a GRB is announced and targeted for follow-up.  If no source exceeds threshold, the trigger system is consulted to see if an improved detector map is available, based on the RAD6000 processing that continued during the DSP's image processing.

\section{Follow-up}

If a GRB is detected and located, the {\it Swift} spacecraft evaluates it for quality and observability, then rapidly slews to make X-ray, optical and UV observations beginning about a minute after the first rate increase detection.  Simultaneously, it sends position and other information down the real-time TDRSS link to allow ground-based follow-up observations.

The BAT also continues its observations.  Light curves in 4 energy bands and up to 128 ms resolution are generated and sent down TDRSS as segments are accumulated, along with the attitude information required to interpret the changing background during the slew.  Event-by-event data, describing every photon detected by BAT from 5 minutes before the GRB to 5 minutes after, is spooled to the spacecraft solid state recorders, for later dumping on the next pass over the Malindi ground station.  The cadence of survey accumulations is accelerated for better time resolution during the afterglow phase.  The GRB is selected as a mask-tagged source for 1.6 s resolution source-specific flux measurements whenever the GRB is in the BAT's field of view.

Over the following days, automatic and ground-planned observations continue until the GRB has faded from sight.


\section{References} { 
\setlength{\parindent}{0pt} 

\hangindent=1em 1. Gehrels, N. {\it et al.}, in {\it Gamma-ray Burst and Afterglow Astronomy 2001}
edited by G.R. Ricker \& R. K. Vanderspek, AIP Conference Proceedings 662, pp. 465-468.

\hangindent=1em 2. Fenimore, E. E.,
Applied Optics {\bf 26} 2760-2769 (1987).

\hangindent=1em 3. Fenimore, E. {\it et al.} , in {\it Gamma-ray Burst and Afterglow Astronomy 2001} edited by G.R. Ricker \& R. K. Vanderspek, AIP
Conference Proceedings 662, pp.  491-493.

}

\bibliography{sample}

\end{document}